\documentclass[12pt]{iopart}
\usepackage{graphicx}

\def\iso#1#2{\mbox{${}^{#2}{\rm #1}$}}
\def\he#1{\iso{He}{#1}}
\def\li#1{\iso{Li}{#1}}
\def\c1#1{\iso{C}{1#1}}
\def\n1#1{\iso{N}{1#1}}
\def\o1#1{\iso{O}{1#1}}

\def\beq{\begin{equation}}
\def\eeq{\end{equation}}
\def\beqar{\begin{eqnarray}}
\def\eeqar{\end{eqnarray}}

\def\la{\mathrel{\mathpalette\fun <}}
\def\ga{\mathrel{\mathpalette\fun >}}
\def\fun#1#2{\lower3.6pt\vbox{\baselineskip0pt\lineskip.9pt
  \ialign{$\mathsurround=0pt#1\hfil##\hfil$\crcr#2\crcr\sim\crcr}}}

\def\msol{\mbox{$M_\odot$}}

\def\mratio{\mu}

\begin{document}

\title{FUSE Deuterium Observations: A Strong Case For Galactic Infall}

\author{Tijana Prodanovi\'{c}}
\address{Department of Physics, University of Novi Sad \\
Trg Dositeja Obradovi\'{c}a 4, 21000 Novi Sad, Serbia}
\ead{prodanvc@if.ns.ac.yu}

\author{Brian D. Fields}
\address{Department of Astronomy, University of Illinois \\
1002 W. Green St., Urbana IL 61801, USA}

\begin{abstract}

Measurements of deuterium  in the local interstellar medium
have revealed large variations
in D/H along different lines of sight. 
Moreover, recent 
{\it Far Ultraviolet Spectroscopic Explorer} (FUSE)
measurements
find D/H to be anticorrelated with
several indicators of dust formation and survival,  
suggesting that interstellar deuterium suffers significant depletion
onto dust grains.
This in turn implies that 
the total deuterium abundance in the local Galactic disk could be
as high as 
$\sim 84 \%$ of the primordial D abundance.
Because deuterium is destroyed in stars, its abundance
in the interstellar medium (ISM) also tells us about the 
fraction of material
which has never been processed through stellar environments.
Therefore, the new report of such high ISM deuterium abundance
implies that {\em most} present-day interstellar 
baryons are {\em unprocessed}.
It was proposed that the infall/accretion of pristine gas
is needed to explain such a high deuterium abundance.
However, we point out that
the infall needed to maintain a high present-day
D/H is, within the preferred models, in tension with observations that 
gas represents only some $\sim 20\%$ of Galactic
baryons, with the balance in stars.
This small gas fraction implies that, integrated over Galactic history,
{\em most} baryons have been 
sequestered into stars and stellar remnants.
We study this tension in the context of
a wide class of Galactic evolution models for
baryonic processing through stars,
which show that deuterium destruction
is strongly and cleanly correlated with the
drop in the gas fraction. 
We find that
FUSE deuterium observations and Galactic gas fraction estimates
can be reconciled in some models;
these demand a significant infall rate
of pristine material that almost completely balances the 
rate of star formation. These successful models also
require that the average fraction of gas that is
returned by dying stars is less than $40 \%$ of the initial stellar mass.
Cosmological implications of dust depletion of D in high-redshift systems
are discussed.

\end{abstract}

\maketitle


\section{Introduction}

Deuterium has a unique nucleosynthetic
property that it is only created in the big bang
nucleosynthesis (BBN) while all other processes, due to deuterium's fragile nature, destroy 
it \cite{els,pf03}. Thus the deuterium abundance should monotonically decrease after BBN.
This gives deuterium a key role in cosmology as a cosmic ``baryometer'' \cite{st}. 
Unlike any other light element, deuterium provides a precise measure
of the primordial abundance but it can also serve as a probe of how much 
gas was processed in stars.

High-redshift ($z \sim 3$) D observations in QSO absorption systems
give a deuterium abundance
$({\rm D/H})_{\rm QSOALS} = (2.82 \pm 0.27) \times 10^{-5}$ \cite{o'meara}, which
is in agreement with the one
inferred \cite{cfo03,cyburt,coc}
from big-bang nucleosynthesis
calculations in concert with the WMAP 
determination of the baryon-to-photon ratio.
This concordance of BBN theory at $z \sim 10^{10}$,
CMB theory and observation at $z \sim 1000$, and
QSO observations at $z \sim 3$ is a strong test of 
the hot big bang and marks a major achievement of the
standard cosmology.

However, measurements of the deuterium abundance in the 
local interstellar medium (ISM) have within a past decade revealed large variations
over different lines of sight (see eg. \cite{jenkins,hebrard,sonneborn}).
Recent FUSE observations \cite{hoopes,linsky} showed that these variations
can be greater than a factor of 3.
In the hot Local Bubble immediately 
surrounding the Sun, 
FUSE observations have very precisely measured a	
deuterium abundance
$({\rm D/H})_{\rm ISM} = (1.56 \pm 0.04) \times 10^{-5}$ 
\cite{wood}.
Over more distant lines of sight
FUSE finds large variations in gas-phase
deuterium, with values scattering far above and below the
Local Bubble measurement; 
\cite{linsky} tabulate values
spanning $({\rm D/H})_{\rm ISM} = (0.5 - 2.2) \times 10^{-5}$.

Linsky {\it et al.} \cite{linsky} have argued that deuterium is 
depleted onto dust
grains efficiently \cite{draine,jura}, and preferentially relative to hydrogen.
In support of this hypothesis, they show that
D/H is significantly anticorrelated with
the depletion of refractory elements like iron and silicon,  
and positively correlated with ${\rm H}_2$ rotational temperature.
These trends imply that D itself is highly refractory,
and that its gas-phase abundance is highest in
environments likely to erode or destroy dust grains.
These trends are also the opposite of naive expectations 
that local nucleosynthetic enrichment of metals by stellar
ejecta should be anticorrelated with deuterium which has
been astrated.

If significant amounts of deuterium 
are indeed locked in dust grains, then
the gas-phase D abundance is only a lower limit
to the total interstellar D inventory.
Linsky {\it et al.} \cite{linsky} argue that the {\it true} 
interstellar deuterium abundance lies at the upper end of the
distribution of gas-phase measurements, and 
estimate 
$({\rm D/H})_{\rm ISM+dust} \geq (2.31 \pm 0.24) \times 10^{-5}$.
This value is consistent with the recent deuterium measurement
in the warm gas of lower Galactic halo where
${\rm D/H} = 2.2^{+0.8}_{-0.6} \times 10^{-5}$ \cite{savage}.
If the interstellar deuterium abundance is indeed as large
as these recent data suggest,
it remains surprisingly close to the primordial deuterium
abundance.

The ratio $({\rm D/H})_{\rm ISM} / ({\rm D/H})_{\rm p}$ 
of the {\em total} (including the potential dust-depleted fraction) 
deuterium abundance in
the {\em present} interstellar medium to the primordial value
has long been recognized as a key probe of Galactic
evolution.  Specifically, this ratio quantifies the
fraction of material in the present-day ISM which has
{\em never} been processed through stars.
For the purpose of our work we will adopt the primordial
deuterium abundance of $({\rm D/H})_{\rm p} = 2.75^{+0.24}_{-0.19} \times 10^{-5}$ 
\cite{cfo03}. This value is consistent, within errors, with other calculations
\cite{cyburt,coc,serpico} but is the highest central value of WMAP-based
evaluations and thus gives the most conservative ISM-to-primordial
deuterium abundance ratio.
For the ISM deuterium abundance we will use
a range of gas-phase deuterium 
$({\rm D/H})_{\rm ISM} = (0.5 - 2.2) \times 10^{-5}$
\cite{linsky} with the central value of $({\rm D/H})_{\rm ISM} = (1.56 \pm 0.04) \times 10^{-5}$ 
\cite{wood}. With this choice of abundances we find that the 
interstellar gas-phase deuterium is a fraction
\beq
\label{eq:Dfrac_g}
\frac{({\rm D/H})_{\rm ISM}}{({\rm D/H})_{\rm p}} = 0.57^{+0.23}_{-0.52}
\eeq
of the primordial value. Note that large errors quoted in \ref{eq:Dfrac_g}
originate mainly from using the entire observed ISM deuterium abundance range.
However, Linsky {\it et al.} \cite{linsky}
FUSE ISM deuterium analysis corrects this fraction upwards to
account for dust depletion, giving
\beq
\label{eq:Dfrac_tot}
\frac{({\rm D/H})_{\rm ISM+dust}}{({\rm D/H})_{\rm p}} = 0.84 \pm 0.09
\eeq
which is significantly higher.  Taken at face value, this
new value suggests that the {\em majority} of baryons in the 
present Galactic ISM
have never been processed through stars!

This new provocative result has important consequences for
the nature of the interstellar medium as well as for
D evolution, and all of its implications have only begun to be tested.
Such tests appeared in \cite{prochaska05,prochaska}, but
with mixed results.  Noting the highly refractory 
nature of titanium, \cite{prochaska05,prochaska} examine the
observed correlation between local Ti/H and D/H abundances.
While a tight correlation is found, 
its comparison with correlations of other refractory elements does not yield 
a consistent picture. Specifically, Fe and Si, which are
less refractory than Ti, show steeper correlation with D/H,
contrary to what is expected \cite{prochaska}.
These results \cite{prochaska} have been reinforced by another recent study \cite{lallement}.
On the other hand, if depletion onto dust was the cause of low deuterium abundances
at some lines of sight, one would also expect 
anticorrelation between D/H and reddening effects along the same line of sight.
This test was done in \cite{steigman}, and again contrary to the expectations, 
they found multiple cases where
low D abundances were observed along some of the lines of sight that
showed the lowest reddening. 

For the purpose of this paper we adopt the Linsky {\it et al.} \cite{linsky} interpretation
that the large scatter in the local deuterium measurements is truly due to
large depletion of deuterium onto dust, and thus examine the consequences of the
resulting true local deuterium abundance that is now close
to its primordial value. 

Though some Galactic chemical evolution (GCE) models
can be consistent with high observed D abundances and the
high value of the true abundance \cite{linsky},
most of them require some continuous infall of primordial gas
that would replenish the ISM deuterium \cite{romano}. 
Moreover, the infall of deuterium-rich material also seems to be needed in order to explain
 high deuterium abundances that were observed in the Galactic Center \cite{lubowich}.
Turning to cosmological scales, both infall (accretion) and outflows appear in 
models of cosmic chemical evolution \cite{pei,pfh}
and in the cosmic context, deuterium astration
has been shown to be sensitive to the details of
gas processing into stars and flows in and out of galaxies \cite{daigne}.
Thus, we reexamine the infall rates needed to explain such high local
deuterium abundance, and moreover, 
analyze the consequences that such result 
bears for high-redshift deuterium observations.
 
In particular, we find that combining the new local deuterium observations
of Linsky {\it et al.} \cite{linsky} with even a wide uncertainty range of measurements of the gas mass fraction
\cite{holmberg2000,holmberg2004,flynn}, 
can be, within a simple GCE model \cite{clayton} and with some adopted
stellar processing return fraction (fraction of stellar mass that is returned to the ISM
in the instantaneous recycling approximation), very constraining for 
the required and allowed infall rates. The tightness of such constraint then only depends
on the assumed return fraction. Specifically, adopting a higher return fraction
value narrows down the infall rate constraint to almost a single value-- star-formation
is almost completely balanced by the infall. On the other hand,
a lower return fraction leaves room for some wider, but still quite constraining, range
of infall rates, where non-zero infall is required. 
Thus, the combination of the two observables, local deuterium abundance measurements of FUSE
and the gas mass fraction, becomes a very powerful and constraining probe of the infall rates.

Finally, in the \ref{sect:metalev} of this paper we also point out a ``universality'' in deuterium evolution
within GCE infall models, where we find that
infall GCE models for deuterium evolution {\em versus metal yields}
show convergent behavior nearly independent
of the rates of primordial gas  infall.

Though the new high estimate of the true local deuterium abundance does intuitively
indicate large infall at play, only in concert with gas mass fraction estimates
does it become a strong constraint of the infall rate, while it on the other hand
does not necessarily require large metal yields. 

\section {Infall and Chemical Evolution of the Milky Way}

\subsection{Formalism}

We construct a model for Galactic evolution with infall.
To characterize the infall rate, we adopt
such parameterization \cite{larson}
in which the gas mass 
infall rate (i.e., baryonic accretion rate)
is arbitrarily proportional to the star formation rate
\beq
\label{eq:infall}
\frac{dM_{\rm baryon}}{dt}=\alpha \psi
\eeq
Here $\psi(t)$ is the star formation rate, while
$\alpha$ is the proportionality constant which characterizes
the strength of infall \cite{larson}. 

The infall prescription of (\ref{eq:infall})
offers (for constant $\alpha$) great mathematical 
convenience, but must be justified on physical grounds.
Models of this sort have been used
to study chemical evolution of the cosmic ensemble of
galaxies \cite{pei}.
Remarkably, the underlying infall/star-formation connection
in these simple heuristic models 
find support in the detailed {\em ab initio}
simulations \cite{keres}
that follow forming galaxies in a 
$\Lambda$CDM universe with gas dynamics
and subgrid prescriptions for star formation and feedback.
These models \cite{keres} find that from $z = 0-5$, the global-averaged
cosmic star formation rate and galactic gas accretion 
rates not only are nearly proportional, but moreover
the rates are nearly identical, i.e., 
in our language their results give
a nearly constant value $\alpha \approx 1$
over this large range in redshift and time.
Observationally, a strong correlation between gas fraction
and present-to-integrated star formation rate,
as measured by the $u-K$ color, has been found \cite{kannappan}. This suggests
that gas availability rapidly leads to, and possibly regulates,
gas consumption via star formation; this is also what is
implied by (\ref{eq:infall}).

In this estimate we will assume that the infalling material
is pristine and that the deuterium abundance in the infalling gas is primordial.
This assumption is supported by observations of low-metallicity high-velocity clouds
where deuterium abundance was measured to be close to primordial \cite{sembach}.
Moreover, due to 
the fragile nature of deuterium we can safely assume that all of the deuterium
is destroyed once it gets in stars.
Thus, following \cite{clayton,tinsley}, we have
\beqar
\label{eq:mrate}
\frac{dM_{\rm ISM}}{dt}=-(1-R)\psi+\alpha \psi    \\
\label{eq:dmrate}
\frac{d}{dt}(DM_{\rm ISM})=-D\psi+D_p \alpha \psi
\eeqar
Here $M_{\rm ISM}$ is the ISM mass (gas+dust) while $R$ is the return fraction,
i.e., the fraction of stellar mass that is returned to the ISM
in the instantaneous recycling approximation. $D$ is
the deuterium mass fraction for a given epoch
$X_{D} \equiv \rho_{\rm D}/\rho_{\rm baryon}$, which we can write as
\beq
\label{eq:mfrac}
D \equiv X_{D} \cong 2 \left( \frac{\rm D}{\rm H} \right) X_{\rm H}
\eeq
where $X_{\rm H}$ is the hydrogen mass fraction.
Primordial deuterium mass fraction is labeled as $D_p$.

Gas accretion through infall and processing through stars 
both change the ISM gas mass.  It is convenient to quantify
these effects via the ratio
$\mratio \equiv  M_{\rm ISM}(t)/M_{\rm baryon,0}$ of gas
at time $t$ to the initial baryonic mass $M_{\rm baryon,0}$. This quantifies
the significance of infall, because when
$\alpha+R > 1$ infall dominates the gas content
of the Galaxy and $\mratio$ increases with time; otherwise,
$\mratio$ decreases with time, the same qualitative trend
as in the closed-box model i.e., the zero-infall limit, in which case
$\mu \rightarrow \omega$ as in (\ref{eq:omega}).
Note also that the ratio 
of {\em total} baryonic mass to initial mass is
\beq
\mu_{\rm tot} \equiv M_{\rm baryon}(t)/M_{\rm baryon,0} 
  = 1 + \alpha \frac{\mu-1}{\alpha + R - 1} 
\eeq
The total integrated accreted mass is thus measured by $\mu_{\rm tot}$, and
one can verify that this expression has $\mu_{\rm tot} > 1$ for all $\alpha > 0$.

Combining (\ref{eq:mrate}) and (\ref{eq:dmrate})
we find the rate of
change of deuterium mass fraction $D$ as
\beq
\label{eq:dm}
M_{\rm ISM}\frac{d{\rm D}}{dM_{\rm ISM}} = \frac{-RD+\alpha(D_p-D)}{\alpha+R-1}
\eeq 
Integrating the above equation in the limits where D mass fraction goes from primordial $D_p$
to present day value $D(t)$, 
we find the solution for deuterium fraction evolution
\beq
\label{eq:dratio}
\frac{D(t)}{D_p} = \frac{R}{\alpha+R}\left(\frac{\alpha}{R}+\mratio^{\frac{\alpha+R}{1-\alpha-R}}\right)
\eeq
We see that the dimensionless
deuterium mass fraction ratio depends only on
the dimensionless gas mass ratio $\mratio$
and thus is insensitive to the absolute Galactic gas mass.

Another observable which directly encodes Galactic baryonic processing
both in star formation and in infall is 
the present gas mass fraction 
$\omega \equiv M_{\rm ISM}(t)/M_{\rm baryon}(t)$.
Combining (\ref{eq:infall}) and (\ref{eq:mrate}) 
we find
\beq
\omega(t) \equiv \frac{M_{\rm ISM}}{M_{\rm baryon}}
 = \frac{1-R-\alpha}{1-R-\alpha \mu(t)} \ \mu(t).
\label{eq:omega}
\eeq
Equations (\ref{eq:dratio}) and (\ref{eq:omega})
relate observables to the model parameters; given a choice of
return fraction $R$ (i.e., a choice of the initial mass function), these two equations
set values for the unknowns $\alpha$ and $\mratio$ (and/or $\mu_{\rm tot}$),
which together respectively quantify the current and integrated rates of
Galactic infall.

The deuterium mass fraction and gas fraction are both excellent probes of
Galactic evolution -- as these observables trace stellar processing
in a particularly clean way.  In our model, both $D/D_p$ and
$\omega$ decrease monotonically with time.
Their declines 
are correlated due to the physics which drives their evolution:
stellar processing
both destroys deuterium and locks up baryons into low-mass stars
and remnants.
We can exploit this correlation to gain insight into
Galactic evolution in general and infall in particular.

At late times, 	
both the deuterium fraction and the gas fraction
approach {\em minimum} values,
regardless of the detailed star formation history.
At late times and for large infall ($\alpha + R > 1$) the parameter
$\mratio^{\frac{\alpha+R}{1-\alpha-R}} \rightarrow 0$,
but this also stays true for 
small infall ($\alpha + R < 1$) scenario since $\mu \rightarrow 0$. 
Therefore, we see that in both cases,
the ratio of deuterium fractions from (\ref{eq:dratio}) goes to 
a minimum value 
\beq
\label{eq:dmin}
\frac{D}{D_p} \rightarrow \frac{D_{\rm min}}{D_p}  = \frac{\alpha}{\alpha+R}
\eeq
which is the ratio of D-inflow to gas-mass-injection rates.
In this same late-time limit, 
if infall is large enough so that $\alpha > 1-R$, 
the gas mass fraction will go to a constant
which we find in the large-$\mu$ limit of (\ref{eq:omega}):
\beq
\label{eq:gasmin}
\omega \rightarrow 
\omega_{\rm min} = 1 - \frac{1-R}{\alpha}
\eeq
i.e., the gas fraction is just given by the ratio of the 
rates of gas (\ref{eq:mrate}) to total (\ref{eq:infall}) mass increase. 
On the other hand, if the infall is small enough
that $\alpha < 1-R$, 
we will have $\mu \rightarrow 0$ and thus $\omega_{\rm min} \rightarrow 0$;
there is no lower limit to the gas faction.

Thus measurements of deuterium and the ISM gas content
each place {\em upper} bounds on infall.
Namely, 
requiring $D_{\rm min} < D_{\rm obs}$ forces 
\beq
\alpha < \frac{D_{\rm obs}/D_p}{1-D_{\rm obs}/D_p} \  R 
\eeq
Similarly, demanding $\omega_{\rm min} > \omega_{\rm obs}$
requires
\beq
\alpha < \frac{1-R}{1-\omega_{\rm obs}}
\eeq
We emphasize that these limits are independent of the
details of star formation history, and free of any uncertainties
in nucleosynthesis yields since metallicity does not enter
at all (and deuterium is totally destroyed in stars).

The minima in (\ref{eq:dmin}) 
and (\ref{eq:gasmin}) teach important lessons for Galactic evolution.
In models without infall, both deuterium and interstellar gas
approach zero as stellar processing becomes large.  
Hence a closed box can only accommodate the high FUSE deuterium
if there is little processing, and thus necessarily a 
high gas fraction.  As we will see, this arrangement is ruled
out in realistic models.
But we see that the presence of 
infall--i.e., nonzero $\alpha$--places nonzero {\em lower} bounds on
deuterium and gas fractions.
Physically, this is of course because
primordial infall replenishes interstellar gas and deuterium.
The presence of infall thus opens the possibility
that the high deuterium abundances seen by FUSE
can be reconciled with significant stellar processing.
But this possibility does not guarantee a solution, as
nonzero infall also raises the gas fraction, which may or may
not remain in agreement with observations.

Thus we see that once a return fraction $R$ has been specified,
one can constrain infall based on deuterium and/or gas observations.
As we will see, using present observations, these limits are 
quite constraining.  Also, the two limits are independent,
and for present observations are competitive with each other.

\subsection{Model Input Parameters and Observational Constraints}

Our goal is to use deuterium and gas fraction observations 
in the context of our model to probe Galactic infall.
To do this, we must specify the observed constraints and their
uncertainties.  In the model, infall is parameterized via
its ratio $\alpha$  to the star formation rate; in addition,
we must also specify the return fraction $R$ 
which too has uncertainties which we must address.

The chemical evolution formalism most naturally uses mass
fraction as an abundance measure.
For this reason, relate the D/H ratio to the
mass fraction $D$ via the definition
(\ref{eq:mfrac}) as
\beq
\frac{D_{\rm ISM}}{D_{\rm p}}= \frac{X_{D, \rm ISM}}{X_{D, \rm p}} =
\frac{X_{H, \rm ISM}}{X_{H, \rm p}} \frac{(D/H)_{\rm ISM}}{(D/H)_{\rm p}}
\eeq
Following this we convert the D/H observational constraints
in (\ref{eq:Dfrac_g}) and (\ref{eq:Dfrac_tot}) to reflect 
corresponding mass fractions
\beqar
\label{eq:Dmfrac_g}
\frac{D_{\rm ISM}}{D_{\rm p}}= 0.53^{+0.22}_{-0.36}    \\
\label{eq:Dmfrac_tot}
\frac{D_{\rm ISM+dust}}{D_{\rm p}}= 0.78 ^{+0.11}_{-0.10}
\eeqar
where we have used the ratio of hydrogen ISM-to-primordial mass fraction 
$X_{\rm H,ISM}/X_{\rm H,p}=0.93$ \cite{ag,cyburt}.

The present Galactic interstellar
gas mass is known to within a factor $\sim 2$ 
\cite{holmberg2000,holmberg2004,flynn}
\beq
\label{eq:gasfrac}
\omega_{\rm obs} \equiv 
  \left( \frac{M_{\rm ISM,MW}}{M_{\rm baryon,MW}} \right)_{\rm obs}
  \sim 0.07-0.30
\eeq
which is typical for similar type spiral galaxies (see eg. \cite{silk07}).
We will see that even with these significant uncertainties,
we can places strong constraints on the
infall rate $\alpha$. 

To estimate
the return fraction $R$, i.e. the mass fraction of stellar population 
that is returned to the ISM, we use the approximation of \cite{iben}
to determine stellar remnant masses $m_{\rm rem}(m)$
and thus to determine the returned mass $m_{\rm ej}(m) = m- m_{\rm rem}(m)$.
If one defines a return fraction $R(m) = m_{\rm ej}(m)/m$ for each
progenitor mass $m$, then the global
return fraction thus represents a mass-weighted average
of this quantity over
the initial mass function (IMF) $ \phi(m) \equiv dN/dm$:
\beq
R = \frac{\int_{m_{\rm L}}^{m_{\rm U}} dm \ R(m) \ m  \ \phi(m)}
         {\int_{m_{\rm L}}^{m_{\rm U}} dm \ m \ \phi(m)}   
\eeq

For our numerical results we adopt
mass ranges $8\le m_{\rm SN}/M_\odot \le 100$ and
$0.8 \le m_{\rm AGB}/M_\odot \le 8$.
Using these with the IMF in the form of \cite{salpeter}
$\phi(m) \propto m^{-2.35}$, we find $R = 0.31$,
which, to the first decimal, does not change with changing mass ranges of a given
stellar population. 
Of course, the return fraction strongly depends on the choice of the IMF. Though widely
used and informative, this IMF \cite{salpeter} is not a realistic one. Adopting a modern and
more realistic IMF \cite{bg,kroupa} that is flatter in the higher-mass regime would
obviously yield a larger return fraction closer to $R \sim 0.4$.
On the other hand, it is still possible to engineer smaller return
fractions.  For example, some very massive stars may
collapse directly to a black hole without first returning
their nucleosynthetic products \cite{bb}. 
This would introduce a cutoff in mass above which there would be no gas return, 
which would in turn result in a lower return fraction.

We will find that for only some values of $R$
is it even  {\em possible} to find
a Galactic evolution model which fits both high present-day
interstellar deuterium and the observed gas fraction.
Thus these observations could provide hints about
the IMF, its mass limits, and the physics of stellar gas
return.  To illustrate the possibilities,
we will present our results for the case of $R=0.3$ but also
for a range of $R \in (0.1,0.5)$. 
As we will see, when solutions are possible, uncertainties
in $R$ allow a wider infall rate range.

\subsection{Model Results}

\begin{figure}[htb]
\includegraphics[scale=0.35]{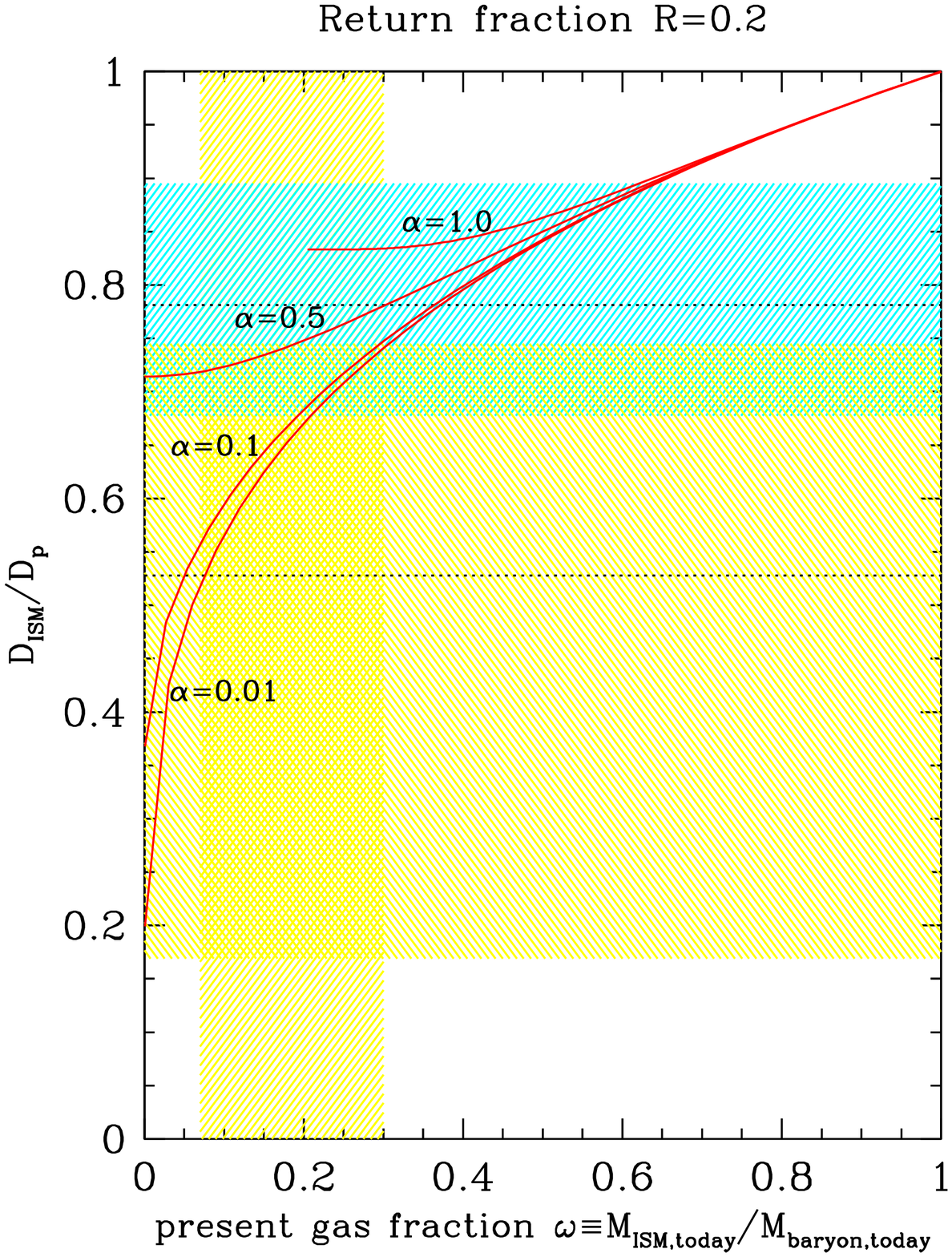}
\includegraphics[scale=0.35]{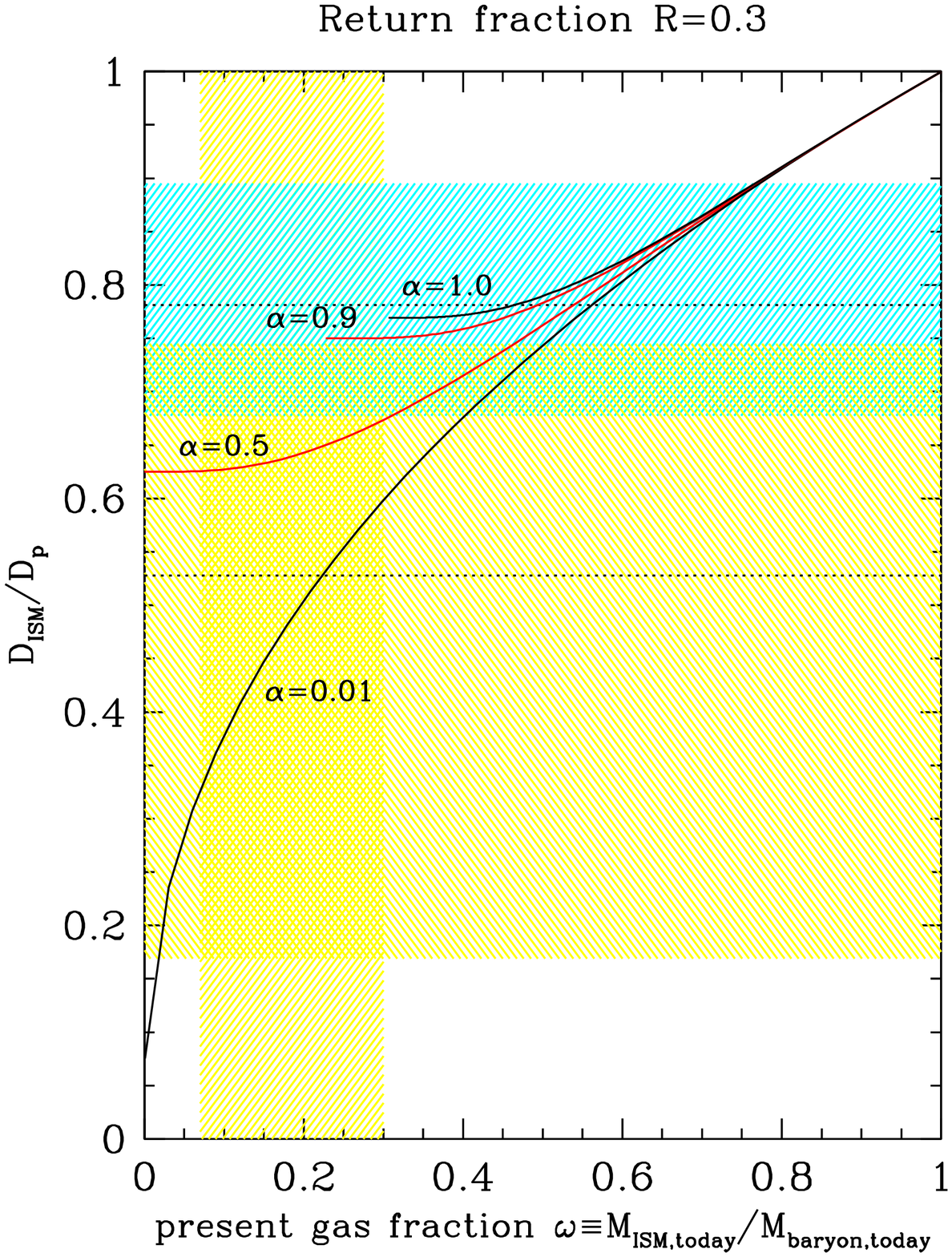}
\caption{
The ratio of the total present-day 
to primordial deuterium mass fraction $D_{\rm ISM}/D_p$ 
as a function of the present gas mass fraction $\omega$.
The two panels reflect different assumed return fractions $R$. 
Dot-dashed black line along with the top (cyan) band corresponds to the ratio of the
deuterium mass fractions for the
ISM deuterium measured by FUSE and corrected for depletion onto dust \cite{linsky} and
primordial deuterium \cite{cyburt} as given in (\ref{eq:Dmfrac_tot}).
Bottom (yellow) band with the central value of (\ref{eq:Dmfrac_g}) \cite{wood} 
given as a dot-dashed black line corresponds to
scatter of deuterium abundances observed along different lines of sight.
Vertical (yellow) band reflects the observed range of present gas mass fraction
$\omega_{\rm obs}$ from
(\ref{eq:gasfrac}).
Different solid curves correspond to infall rates 
with proportionality constant: 
$\alpha=0.01,0.1,0.5,1.0$ left panel, and 
$\alpha=0.01,0.5,0.9,1.0$ right panel.}
\label{fig:Dvsgasf}
\end{figure}

Presented on Figure \ref{fig:Dvsgasf} is the ISM-to-primordial deuterium ratio as
a function of the present gas mass fraction for two different return fractions.
Different curves on the same panel reflect different infall rates. 
The two bands in the $D_{\rm ISM}/D_p$ reflect the ratio of FUSE measured 
$D_{\rm ISM}$ with and without accounting for the possible dust depletion to the
primordial value. These correspond to
top (cyan) band, which includes corrections for a severe dust depletion  
and the central value of 
(\ref{eq:Dmfrac_tot}) \cite{linsky}, and 
bottom (yellow) band with the central value 
given by gas-phase abundances only
(\ref{eq:Dmfrac_g}).
The width of the lower band was dictated by
the scatter of deuterium ISM measurements where the limits came from
the highest and the lowest observed deuterium abundances in the ISM \cite{linsky}. 
The adopted primordial deuterium abundance is that of Cyburt {\it et al.} \cite{cfo03}. 
The vertical (yellow) band in the present gas mass fraction
reflects the observed range of values
for $\omega_{\rm obs}$
in (\ref{eq:gasfrac}) depending on the adopted model
(within errors). 

Figure \ref{fig:Dvsgasf} shows that when the gas fraction is  large,
the rate of infall is not important
and thus all curves that reflect different infall rate $\alpha$ converge. However, in the low gas mass
fraction range which is the one consistent with observations, we see that different
infall rate curves behave very differently. Thus we see that the overlap region between the
observed $\omega$ and the ratio of the new deuterium FUSE measurement that corrects for the dust depletion (top band) 
to the primordial value, can be quite constraining of the infall rate.
Specifically, from the left panel ($R=0.2$) we see that 
all infall rates that are lower than the star formation rate, i.e. $\alpha \la 1.0$,
can be consistent with the measured mass fraction range.
On the other hand, adopting a higher
return fraction of $R=0.3$, as presented on the right panel, is more constraining, and
it allows for only $0.5 \la \alpha \la 1.0$.
We note that the infall rate that is comparable to the
star-formation rate is consistent with the
cosmic-averaged accretion rate found in
\cite{keres}.

\begin{figure}[htb]
\includegraphics[scale=0.5]{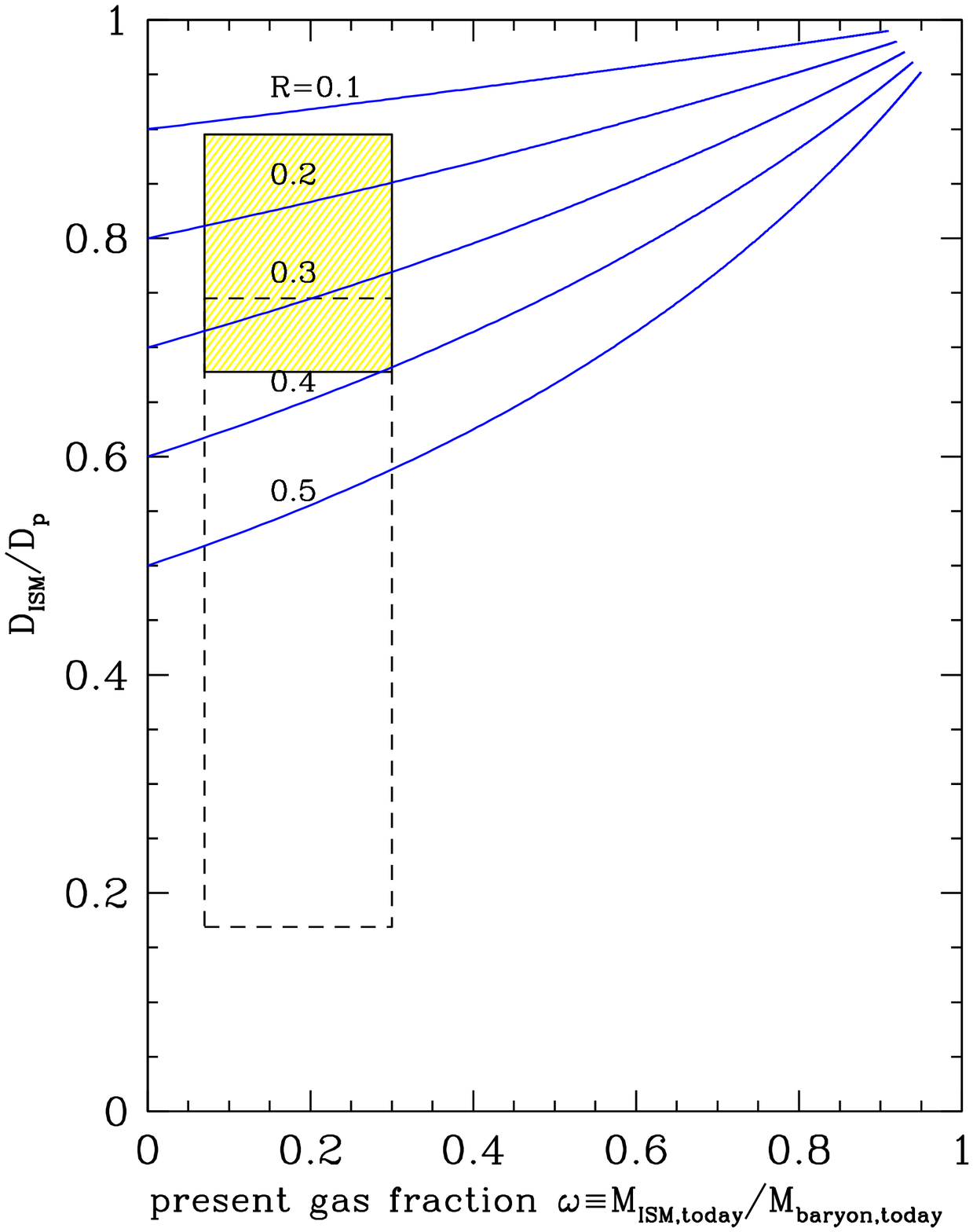}
\caption{The ratio of the {\it limiting} present-to-primordial 
deuterium mass fraction $D_{\rm ISM}/D_p$ 
as a function of the {\it limiting} gas mass fraction $\omega$.
The boxes present the overlap regions between the observed range the of 
present gas mass fraction
$\omega_{\rm obs}$ (\ref{eq:gasfrac})
and measured ISM deuterium abundances with (top, yellow box) and without
(bottom, dashed box) depletion onto dust \cite{linsky}. 
Different solid curves correspond to return fractions labeled.}
\label{fig:envelope}
\end{figure}

It is also apparent in Figure~\ref{fig:Dvsgasf} that
as the infall rate $\alpha$ gets larger the curves that
reflect possible solutions begin at higher $\omega$ e.g.
for $R=0.3$ (Figure \ref{fig:Dvsgasf} right panel), 
$\alpha=0.9$ curve starts at $\omega \approx 0.2$ while
 $\alpha=1.0$ curve starts at $\omega \approx 0.3$.
Thus if one was to plot $D_{\rm ISM}/D_p$ as a function of $\omega$
where $R$ is set, but do that for a wide and continuous
range of infall parameters $\alpha$ and not just for a few discrete values that we used
for the purpose of demonstration, one would find that as the $\alpha$ increased the curves
would move upwards -- to higher $D_{\rm ISM}/D_p$ values, and to the right -- 
would begin at a higher $\omega$. This means that there would be an area in Figure~\ref{fig:Dvsgasf}
where no solutions could be found. This boundary would be defined by combining all starting 
(lowest $\omega$) values for all possible infall parameters $\alpha$, like an envelope,
a limiting curve, above which no solutions can be found, for a given return fraction $R$.

Expressing the infall parameter $\alpha$ from (\ref{eq:gasmin}) and combining it with
(\ref{eq:dmin}) we find the limiting curve in the form of
\beq
D_{\rm min}/ D_p=\frac{1}{1+ R(1-\omega)/(1-R)}
\eeq

On Figure~\ref{fig:envelope} we plot these enveloping curves
for a range of return fractions. The boxes present the overlap region between
gas fraction and observed ISM deuterium measurements -- yellow box corresponds to
dust corrected local deuterium fraction, while dashed box reflects standard local
D observations without any dust depletion being assumed \cite{linsky}. 
This plot further demonstrates the constraining power of
combining deuterium and the gas mass fraction observations-- 
high ($R >0.4$) return fractions
are completely excluded, while low $0.1< R \la 0.4$ can allow for a wider range of infall
rates. 

\begin{figure}[htb]
\includegraphics[scale=0.5]{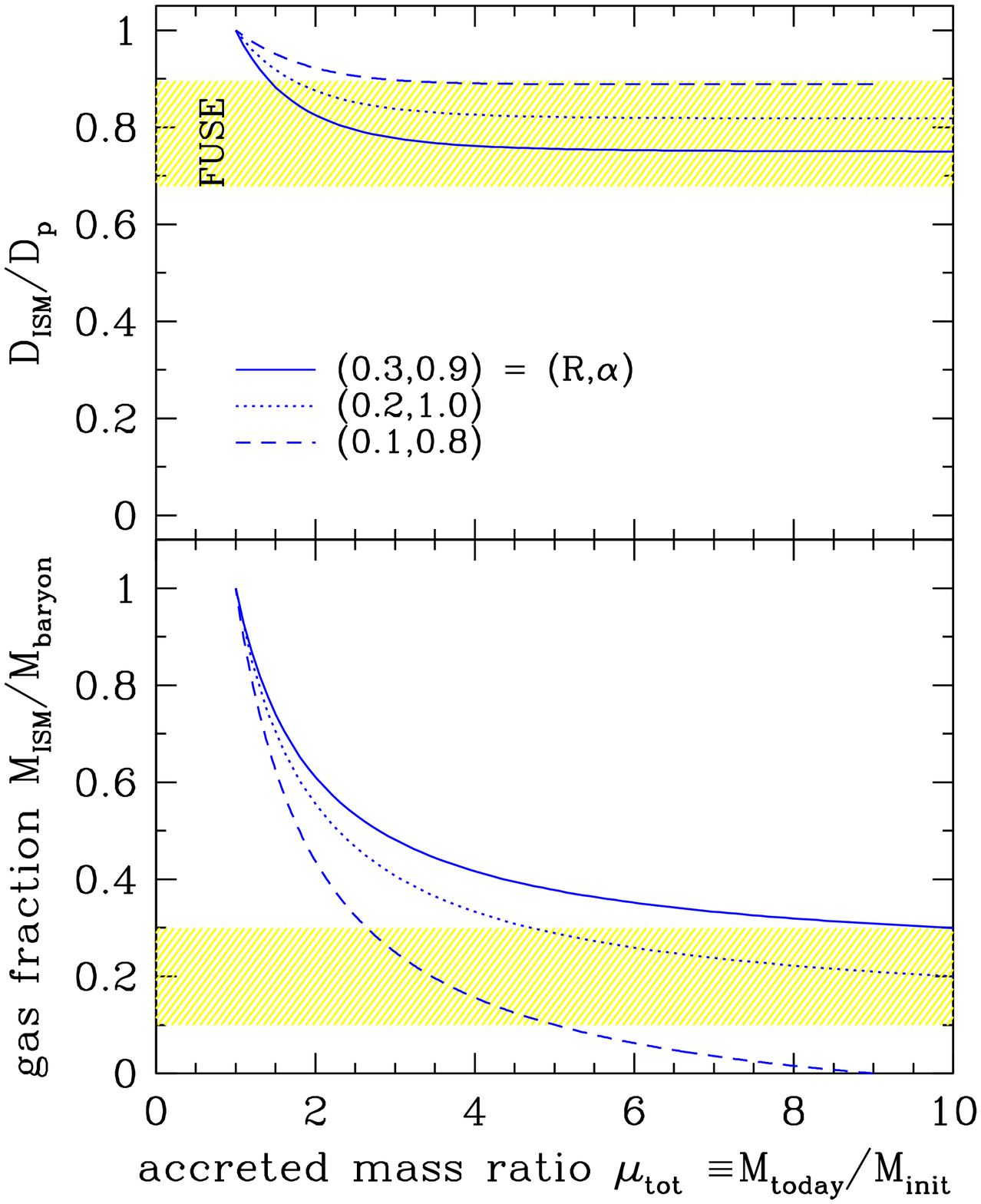}
\caption{The ISM-to-primordial deuterium fraction (top panel)
and gas fraction $\omega$ (bottom panel) as functions of the present total-to-initial baryonic
mass ratio $\mu_{\rm tot}$. The ratio of the total present mass to the total initial mass
is effectively the total accreted mass fraction when no outflow is considered. Yellow bands on both
panels represent the observational requirements and thus the results need to be within them.
As an illustration, three sets of values for model parameters return fraction $R$ and infall rate 
proportionality constant $\alpha$ have been considered: solid curve $R=0.3$, $\alpha=0.9$;  
dotted curve $R=0.2$, $\alpha=1.0$; dashed curve $R=0.1$, $\alpha=0.8$.}
\label{fig:accretion}
\end{figure}

Figure \ref{fig:accretion}
shows the ratio of deuterium fractions $D_{\rm ISM}/D_p$ (top panel) and the gas fraction $\omega$
(bottom panel)
as functions of the ratio of the total present mass to the total initial mass
$\mu_{\rm tot}$, which is in the case where no outflow is considered,
a measure of the total accreted mass, i.e. the accreted fraction. 
The curves correspond to three pairs $(R,\alpha)$ of 
return fractions and infall proportionalities, while
(yellow) bands mark the range of observed values of $D_{\rm ISM}/D_p$
and $\omega$. These parameter pairs were chosen as representative of
models which simultaneously satisfy both observational
constraints. It is obvious that gas fraction, as a function of present-to-initial mass
(bottom panel), is far more constraining, and that constraints on $\mu_{\rm tot}$
for a given set of $(R,\alpha)$ will also fit within the deuterium observational constraint 
(top panel). It is clear that $\mu_{\rm tot} \gg 1$ in all scenarios, which
immediately tell us that accretion is an important factor. 
From this plot we thus see that
the most modest accreted fraction
$2.7 \la \mu_{\rm tot} \la 5$ is allowed in the case of ($R=0.1, \alpha=0.8)$, while
higher return fractions demand $\mu_{\rm tot} \gg 1$, specifically
$\mu_{\rm tot} \ga 10$ for $(R=0.3,\alpha=0.9)$. Since higher return fractions
are preferred, this is in agreement with the result that the {\em most} 
of the present-day Galaxy was assembled by accretion rather than mergers \cite{murali}.

\section{Cosmological Implications of FUSE:  D as a Baryometer}

Up till now, analysis of high-$z$ D/H observations
have assumed that deuterium is not depleted
into dust in the unresolved Lyman-limit and damped Ly-alpha
systems where it is observed.  The FUSE results 
and their interpretation by Linsky {\it et al.} \cite{linsky}
at minimum demand that we 
consider the possibility that this assumption may not be valid.  

The possibility of D depletion onto grains
would imply that D gas-phase abundances probed in QSOALS
represent  {\em lower limit} to the total
D/H.  This effect goes in the same direction as 
astration, but
astration is expected to be a very small correction
in the low-metallicity QSOALS.  On the other hand,
if D dispersion in QSOALS were present due to depletion onto
dust grains, and were as large as suggested by the local Galactic
FUSE  results, the possible upward corrections to the high-redshift
D/H could be significant. 

As with the local FUSE data, the highest
reliable high-$z$ D/H value
would be the best estimate of the true D/H value
and thus the primordial abundance.
One could make this approach more sophisticated
by taking a weighted average of several 
D/H upper limits, in a manner similar to
non-parametric approaches which have been used to
infer the primordial \he4 abundance
\cite{hogan}.
At the moment, the D data are sparse,
and already lower-quality (and lower abundance)
results from Lyman limit systems have been
excluded from the analysis, effectively performing
the upper limit analysis we are suggesting here.

Our other conclusion--that D astration is more or less
independent of infall at a fixed metallicity--also
has implications for high-$z$ deuterium measurements.
Metallicities are available for QSOALS and are small, 
typically $Z \sim 1\% Z_\odot$.  For any infall model 
this implies that $D/D_p \approx 1$ to
within $\sim Z/Z_\odot \sim 1\%$
(via ~\ref{eq:dz2}).
Thus, our results
strengthen the case for using high-$z$ observations 
where metallicity is small.  The caveat is that there
remain possibilities of outflows, which could be metal-rich
and then would not be covered in our analysis.

It is also important to bear in mind that 
empirically, the low-metallicity QSO absorbers
do not show evidence for dramatic 
depletion of metals onto dust grains
\cite{pettini94,pettini97}.
This result stems from the higher ratio of
refractory-to-volatile elements in low-metallicity QSOALS
relative to the local ISM ratio.
The apparent lack of depletion of heavy element onto to
grains  would in turn suggest that D depletion onto grains 
may also be small at high redshifts.

All of these issues leave the situation unsettled, but
several observational tests suggest themselves.  
In light of the local FUSE data,
additional high-$z$ D measurements become all the more valuable.
A larger data set would allow a statistically robust search for
a ceiling which would represent the primordial abundance.  Also
one would look for the kinds of correlations and anticorrelations
with metals and temperature that leads to the dust interpretation
of the FUSE results.  The prescient work of Timmes {\it et al.}
\cite{timmes}
argued that measurements of D ratios with nitrogen and oxygen
in QSOALS would be a very useful check  on chemical evolution
and observationally would eliminate uncertainties due to
ionization corrections.  In particular, they argue that
D/N and D/O can be determined precisely since their atomic properties
dictate that these species are locked to hydrogen and thus
represent the true abundance ratios.
Observation and analysis of the redshift history of D-to-metal ratios 
in quasar absorbers thus becomes a high priority, not only
to determine effects of chemical evolution but also to
detect and quantify the extent of depletion onto dust grains. 

\section{Discussion}

Other work has noted a correlation
between deuterium and gas fraction evolution
which sheds light on our results and the approximations
we have adopted.
Vangioni-Flam {\it et al.} \cite{vop} made a detailed study of the Galactic
evolution of D/H in closed-box models,
and emphasized the strong
correlation of D/H with gas fraction.
Our results confirm this trend for infall models
and for general star formation histories.
These authors \cite{vop} also explored the effect of the instantaneous
recycling approximation (IRA) which we have adopted in this work.
For some of the models (i.e., for some star-formation histories)
they examined, the evolution
{\em with time} of both D/H and gas fraction
are not strongly affected by the IRA.
Models, in which
the time evolution of D/H and gas fraction
can be significantly different
with and without IRA, were also identified \cite{vop}; however, these models lead to unphysically
low D/H at late times.  Moreover, it was found \cite{vop} that 
the evolution of D/H versus gas fraction--i.e.,
the co-evolution of the observables--is most sensitive to the IRA
at intermediate times when
the fractions lie far above observed values;
at gas fraction value close to those observed,
the IRA result for D/H comes to within $\sim 30\%$ of the non-IRA value.
The upshot for our work here--which uses the IRA but allows
for infall and arbitrary star-formation rates--is
that the quantitative constraints on infall are probably 
good to no better than $\sim 40\%$ and that the true results
{\em will} depend on the
details of the Galactic star formation history.
However, our main qualitative point remains unchanged:
a high interstellar D today together with 
a low gas fraction requires large infall.

Although this goes beyond the scope of our simple but instructive model, we will
here briefly comment on two major approximations used in our model --
infall proportional to the star-formation rate, and the lack of outflow. 
If the infall rate is decoupled from the star-formation rate, but in such a way
that there is a large infall very early on, this extreme scenario would amount to
just having a larger initial total mass which would not affect the 
late-time deuterium history probed by ISM observations today.
In a more general framework, one could view our resulting infall rates 
as the average of the true infall history weighted by the
star-formation rate history. Detailed models of deuterium 
evolution with different infall histories can be found in e.g. \cite{steigman92,tosi98}.
As we have seen from Figure~\ref{fig:envelope} the return fraction $R \approx 0.4$, which would
result from using currently preferred IMF \cite{bg,kroupa}, is just barely allowed. However,
if the outflow was included in our model, but in such a way that there is still a positive net
inflow of gas, the resulting curves presented on  Figure~\ref{fig:envelope} would move upwards
on the lower gas fraction side of the plot, allowing for larger return fractions. 
Looking at the Figure \ref{fig:Dvsgasf}, this also means that a given deuterium fraction
could now be facilitated with a smaller infall rate, for a given gas fraction, than in the
no-outflow scenario. This is easily understood when one recalls that the infalling material
is D-rich while the outflowing gas is D-poor, and thus the ISM deuterium is more easily replenished. 
On the other hand, any outflow must also dilute the ISM of metals, which could
result in a demand for unreasonably large stellar metal yields 
(see \ref{sect:metalev} for further discussion).

The evolution of D has been illustrated in
global models of {\em cosmic} chemical evolution \cite{daigne},
with noticing the resulting dependence on gas fraction.
The authors \cite{daigne} note a connection with local interstellar
D/H variations and point out that
local deuterium measurements may not be reliable 
probes of the primordial abundance.
Our findings are in agreement with these general conclusions,
though if deuterium variations are largely due to
dust depletion effects, the difficulties in
using local D/H for cosmic probes stem at least
as much from uncertainties in estimating the
degree of depletion.  
Moreover, this in turn raises the question of
how deuterium dust depletion might affect systems at high
redshift.

Finally, we reiterate that our results have been designed to
explore the consequences of 
proposal that interstellar D/H is quite high
due to a significant depletion of deuterium onto dust grains \cite{linsky} .
That we are able to find chemical evolution models which 
can accommodate this result provides a sort of existence proof
that workable evolutionary scenarios can be found. But in models
which reproduce the high D/H today, the needed return
fractions and metal yields strain at current favored IMF
choices, and suggest both a relatively large sequestration of baryons into
low-mass stars, but also a relatively large fraction of mass
in high-mass stars.
More detailed Galactic chemical evolution studies,
along the lines of \cite{romano} and \cite{steigman},
will be very useful in determining how a
holistic picture of Galactic chemical evolution
can be understood in light of the FUSE data.

\section{Conclusions}

Linsky {\it et al.} \cite{linsky} have used FUSE data to argue that a substantial portion of
the deuterium in the
the local ISM is not in the gas phase but depleted onto dust grains.
This in turn demands upward corrections to the 
local D abundances, placing the true ISM D abundance at a much higher level than standard GCE 
models would suggest. 
In order to explain such high ISM D abundance, a significant infall/accretion of pristine
gas appears to be necessary. 

In this paper we have shown that another
observable highly sensitive to infall
is the gas mass fraction, i.e. the ratio of the mass contained in the gas and the total baryonic mass
at a given time. 
Moreover, both the D and gas fraction increase with larger
infall (primordial in composition).
Within a simple model \cite{clayton} and by assuming
an infall that is proportional to the star-formation rate with
proportionality constant $\alpha$ \cite{larson} our results can be summarized
in the following 
\begin{itemize}
\item
We find a deuterium mass fraction
as a function of the gas mass fraction where the infall rate and the
stellar processing return fraction are input parameters.
This result is plotted on Figure \ref{fig:Dvsgasf} for a range of infall rates.
\item
We constrain the allowed infall rates within the observed ranges of
deuterium and gas mass fractions that are also plotted on 
Figure \ref{fig:Dvsgasf}.
\item
Our constraint
gets stronger with a higher return fraction. The return fraction of $R=0.3$ 
requires a {\it significant} infall/star formation scaling that is in the range of $0.5 \la \alpha \la 1$.
\item
We find the allowed range of return fractions to be $0.1 \la R < 0.4$; see Figure \ref{fig:envelope}.
\item
High return fractions $R>0.4$ that are completely excluded or just marginally allowed 
are preferred by the modern IMF models \cite{bg,kroupa}.
\item
Turning the problem around, recent numerical
simulations {\em predict} $\alpha \approx 1$ \cite{keres},
which would {\em demand} high deuterium today:
$D/D_p \ga 0.77 $ for $R \le 0.3$, while higher return fractions
are completely excluded.\footnote{
We thank the anonymous referee for pointing this out.
}
\item
We find a simple universality -- ISM D fractions for different infall rates all follow 
the same trend as a function of metal yields (see the appendix). This is presented on Figure \ref{fig:DvsZ}. 
Thus high true ISM abundance can be made consistent with standard GCE models with or without 
primordial gas infall.
\end{itemize}

Our results focus on Galactic chemical
evolution, but must ultimately be placed in a larger
context of cosmological evolution \cite{pei,pfh,daigne}.
Our need for infall is broadly consistent with 
hierarchical assembly of galaxies by accretion of satellites.
Indeed, if just 10\% of the $\sim 10^{11} \msol$ baryonic mass of the Galaxy 
is assembled by accretion (as opposed to merging), then the
average infall rate over cosmic timescales comes to $\sim 1 \msol/{\rm yr}$,
i.e., close to the current star formation rate.
Moreover, the range of infall parameters $\alpha \sim 0.5-1.0$  in our
allowed models are in good quantitative agreement with the
cosmic-average results from the gas dynamic simulations \cite{keres}.

Finally, we note here that the Linsky {\it et al.} \cite{linsky} analysis of Galactic deuterium 
raises a larger question: if the local D is so severely
depleted onto dust, should one worry that a similar effect might be at play
in high-redshift QSO D observations? Standard BBN theory can only allow for
a small upward dust depletion corrections to the high-redshift deuterium abundances before
it runs into a problem. Note here however that requiring a higher primordial D abundance
would result in a lower primordial \li7 abundance which could bring resolution to
the pressing disagreement between the observed Spite plateau and theoretical predictions
of the primordial lithium abundance \cite{cfo03}. 
High-redshift studies have found that at redshift $z\sim 2$ dust-to-gas ratio is at the level $\sim 10\%$
of the one measured in our Galaxy \cite{pettini94}. Thus one should not expect as large dust depletion
at high redshifts as has been suggested for the local ISM. 
Nevertheless, the potential problem
that deuterium depletion onto dust could lead to, makes this an important question
to ask in the light of the new true ISM D abundance determination \cite{linsky}. 
A redshift history of the ratio of deuterium-to-refractory elements like Ti and Si \cite{draine,prochaska},
or the deuterium-to-oxygen ratio \cite{timmes}
would be essential for settling the question of the level of D depletion onto dust
at high redshifts.      

\ack
We are grateful to Gary Steigman for insightful comments.
We would also like to thank the referee on his comments that helped make this paper better.
BDF thanks Shelia Kannappan for enlightening discussions. 
The work of TP is supported in part by the Provincial Secretariat for Science and 
Technological Development, and by
the Republic of Serbia under project number 141002B.

\appendix

\section{Metallicity Evolution}

\label{sect:metalev}

A more accessible but more model-dependent
observable of our local chemical evolution is metallicity, 
whose evolution is normally the focus of GCE models.
Although somewhat lower, we will approximate the metallicity of the local ISM 
with a solar value $Z \approx Z_\odot$ (see eg. \cite{wilms} and references therein). However, in primordial infall models, stellar enrichment of
the ISM is partially diluted by the infall, thus 
one has to ask what it takes to achieve a solar metallicity ISM.

Taking $Z$ to be the interstellar metal mass fraction, while $Z_{\rm ej}$ is the
ejecta metal mass fraction (the metal mass that stars return to the gas) 
we have
\beq
\frac{d}{dt}({\rm Z}M_{\rm ISM})=-{\rm Z}\psi+{\rm Z_{\rm ej}}R \psi
\eeq
where we assume the infall of pristine material with no metals. This, now, similar to (\ref{eq:dm}) gives
\beq
\label{eq:zm}
M_{\rm ISM}\frac{d{\rm Z}}{M_{\rm ISM}}= \frac{-Z\alpha+R(Z_{\rm ej}-Z)}{\alpha+R-1} 
\eeq

For the purposes of our analytic treatment,
we can approximate metal yields in ejecta in two ways.

(a) {\em Metallicity-Independent Yields}.
We can assume a constant yield:
$Z_{\rm ej}=Z_{\rm synt}=const$, i.e., {\it all} stars make new metals with constant yields
\footnote{We note here that in our notation, $Z_{\rm synt}$ represents a {\it true yield} of metals
rather than the effective yield which is commonly used 
in the literature \cite{tinsley}
and is defined as $y_Z = Z_{\rm synt}/(1-R)$. }
which are independent of time and of
inherited metallicity.
Physically, this ignores the recycling of unprocessed initial 
metals retained in stellar envelopes and returned to the
ISM upon stellar death.  This is a negligible effect for
massive stars which will die as metal-rich supernovae,
but can be important for lower-mass AGB stars.
Thus this yield approximation can be viewed as an indication of
the effect of the lock-up of metals in AGB stars of
non-negligible lifetime; in this sense, this yield approximation
offers a check on the impact of the instantaneous recycling approximation
on metallicity evolution.

(b)
{\em Recycled-Inheritance Yields}.
 Alternatively, we can account for initial stellar metallicity
by writing
$Z_{\rm ej}(t)=Z(t)+Z_{\rm synt}$, where now $Z_{\rm synt}=const$ represents
the net {\em new} metal mass fraction produced by a stellar generation, while
$Z(t)$ is the metal mass fraction that a star was born with.
Thus, pre-existing metals from prior generations are returned as
well; in the instantaneous recycling approximation
we have $Z(t_{born})=Z(t_{end}) \equiv Z(t)$.
The return of initial metal mass
is unimportant for massive stars, which produce far more
metals than they are born with. However, the effect is large for
lower-mass stars, which represent the majority of the returned mass, and
which do not make significant amounts of new metals.
In this scenario the yields are manifestly 
metallicity dependent, 
in a way that it becomes important when the ISM metal mass fraction
$Z$ approaches $Z_{\rm synt}$; at $Z \ll Z_{\rm synt}$ the difference between the two
approximations is negligible.

In the constant-yield approximation (a), by defining new variables $\Delta(t) \equiv D(t)/D_p$
and $\zeta(t) \equiv Z(t)/Z_{\rm synt}$, we can rewrite(\ref{eq:dm}) and (\ref{eq:zm}) as
\beqar
\label{eq:delta}
\dot{\Delta}M_{\rm ISM} &=& -R\Delta +\alpha(1-\Delta)    \\
\label{eq:zeta}
\dot{\zeta}M_{\rm ISM} &=& -\alpha \zeta+R(1-\zeta)
\eeqar
If one makes a further change of variables $\zeta=1-u$,
then we see that $u$ has an evolution equation identical to that for
$\Delta$ (i.e., $u$ and $\Delta$ both must satisfy ~\ref{eq:delta}).
Since $\Delta$ and $u$ both have the same initial value $\Delta(0) = u(0) = 1$,
and satisfy the same equations, then we have $u = \Delta$ for all $t$, 
which then means $\zeta=1-\Delta$ for all $t$. 

This means that there is a simple connection
between deuterium mass fraction and the stellar yield of metals that is
independent of infall rate, and is in the form of
\beq
\label{eq:universal1}
\frac{D(t)}{D_p} = 1-\frac{Z(t)}{Z_{\rm synt}}
\eeq

\begin{figure}[ht]
\includegraphics[scale=0.35]{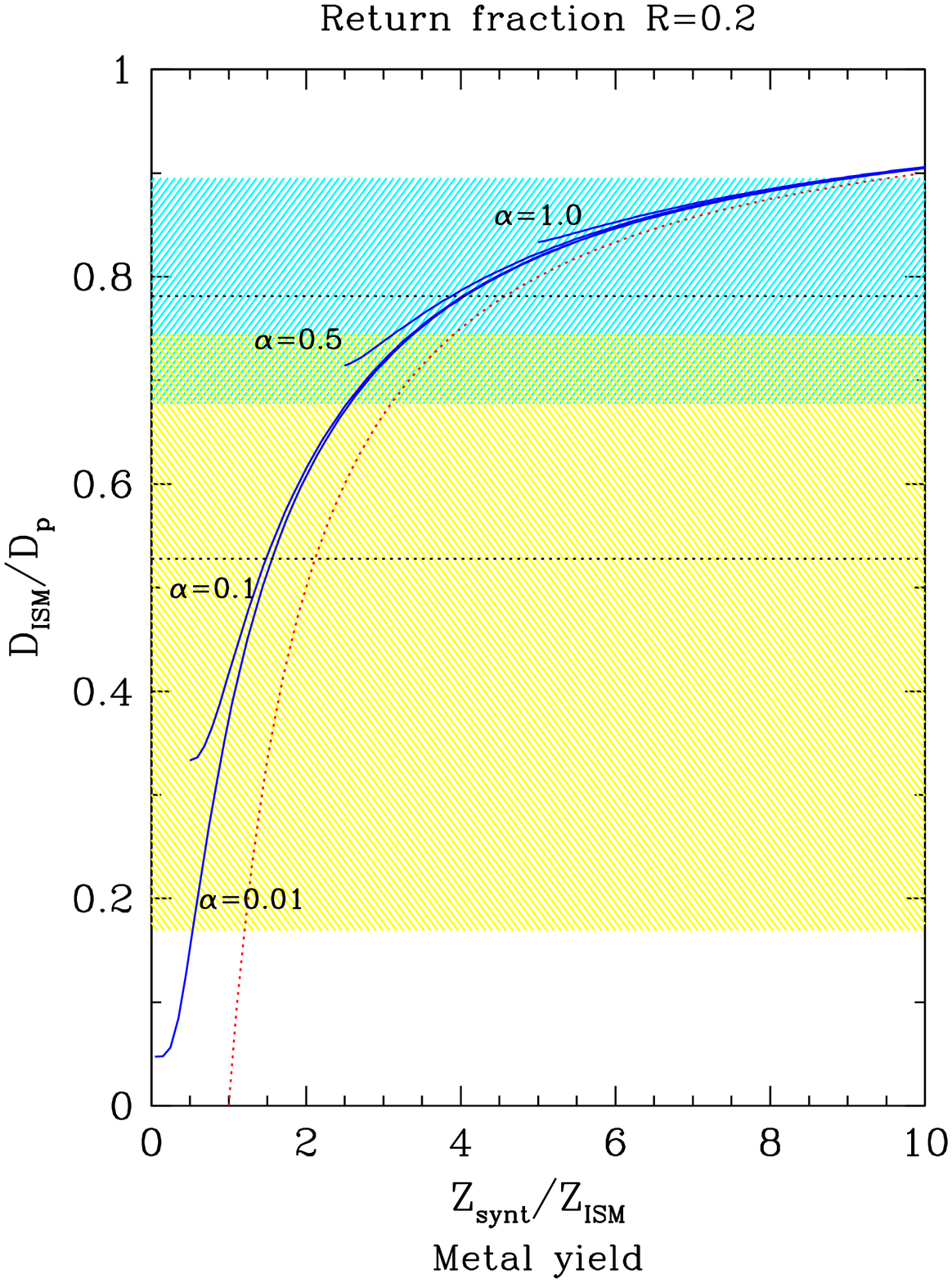}
\includegraphics[scale=0.35]{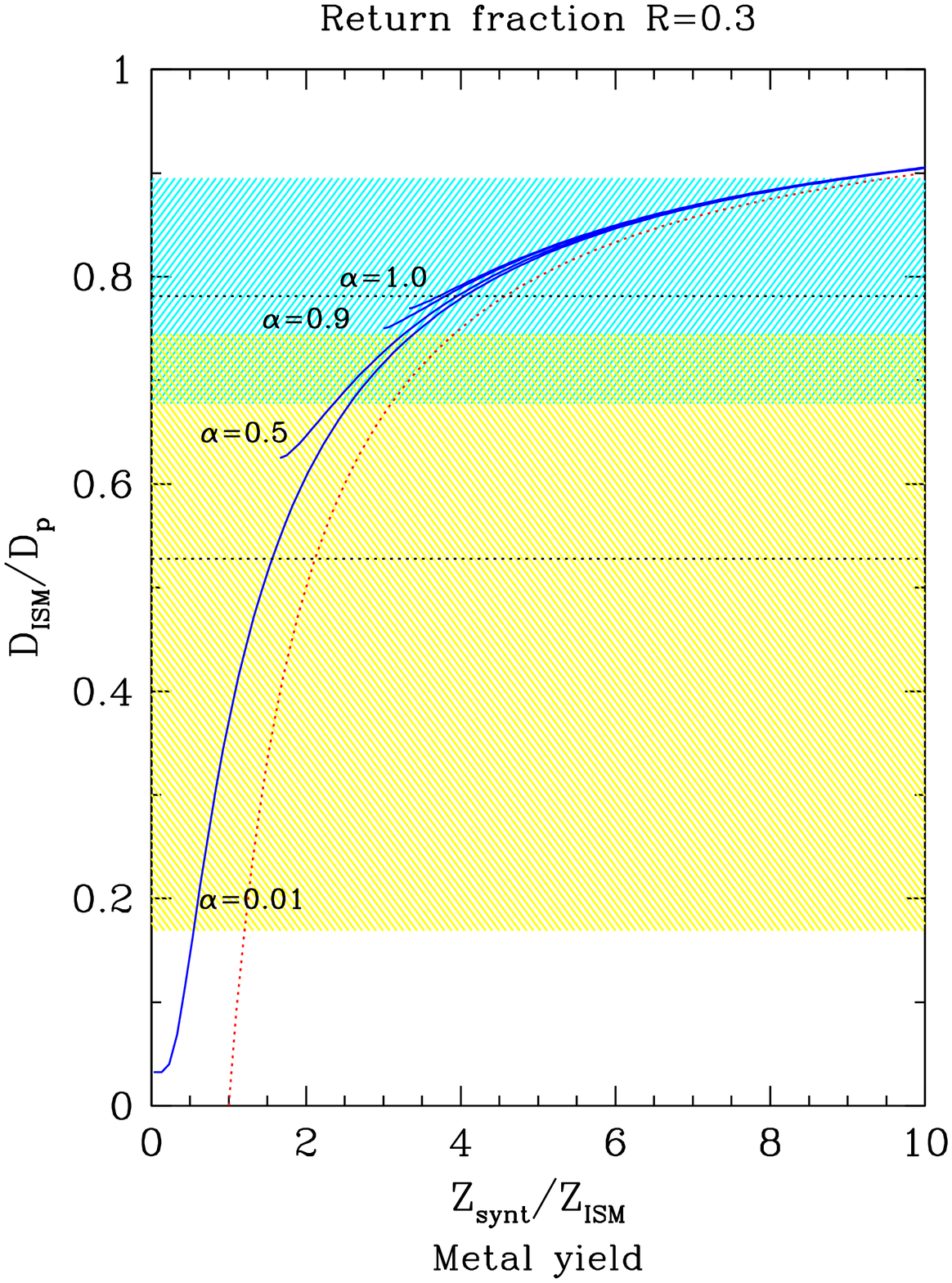}
\caption{
The ratio of the total present-day 
to primordial deuterium mass fraction $D_{\rm ISM}/D_p$ 
as a function of $Z_{\rm synt}/Z_{\rm ISM}$, i.e., the metal yield at fixed ISM metallicity. 
The two panels reflect different assumed return fractions $R$. 
The dotted curve (red) reflects an extreme case where only high mass stars 
are taken into account, i.e., $Z_{\rm ej}=Z_{\rm synt}=const$.
Solid curves (blue) reflect a model where $Z_{\rm ej}(t)=Z(t)+Z_{\rm synt}$, i.e.
where both high mass (give constant metal yields $Z_{\rm synt}$) 
and low mass stars (return initial metal abundance $Z(t)$) are taken into account.
Different solid curves correspond to infall rates 
with proportionality constant: 
$\alpha=0.01,0.1,0.5,1.0$ left panel, and 
$\alpha=0.01,0.5,0.9,1.0$ right panel.
Dot-dashed black line along with the top (cyan) band corresponds to the ratio of the
ISM deuterium measured by FUSE and corrected for depletion onto dust and
\cite{linsky}
primordial deuterium \cite{cyburt} as given in (\ref{eq:Dmfrac_tot}).
Bottom (yellow) band with the central value of (\ref{eq:Dmfrac_g}) \cite{wood} 
given as a dot-dashed black line corresponds to
scatter of deuterium abundances observed along different lines of sight.
\cite{linsky}.}
\label{fig:DvsZ}
\end{figure}

The result of this limiting case is plotted as a dotted (red) curve
on Figure \ref{fig:DvsZ}.\footnote{
Our discussion has focused on the illustrative case 
in which infall
has the form $f = \alpha \psi$.
However, one can show quite generally that
(~\ref{eq:universal1}) holds for arbitrary infall rates
within the instantaneous recycling approximation.
}

In the metallicity-dependent yield approximation (b),
(\ref{eq:zm}) now becomes
\beq
M_{\rm ISM}\frac{d{\rm Z}}{M_{\rm ISM}}= \frac{-Z\alpha+RZ_{\rm synt}}{\alpha+R-1} 
\eeq
which gives the solution in the form
\beqar
\nonumber
\frac{1-R-\alpha}{\alpha}\ln \frac{-{\rm Z}
 \alpha+{\rm Z_{\rm synt}}R}{{\rm Z_{\rm synt}}R} &=&
\ln \mratio  \\
\label{eq:zratio2}
Z(t)/Z_{\rm synt} &=& \frac{R}{\alpha} \left( 1-\mratio^{\frac{\alpha}{1-\alpha-R}} \right)
\eeqar
At late times, $\mratio^{\frac{\alpha}{1-\alpha-R}} \rightarrow 0$
for all nonzero $\alpha$, and we see that in the presence of infall,
the ISM metal mass fraction goes to 
a limiting maximum value 
\beq
\label{eq:zmax}
Z_{\rm ISM} \rightarrow Z_{\rm max} = \frac{R}{\alpha} Z_{\rm synt}
\eeq
We see that infall sets an upper limit to metal mass fraction, and
thus the observed ISM metallicity can set limits on
infall, given estimates of the mean stellar yield.
These limits are comparable to those from D and gas fraction measurements,
but are more model-dependent in that they depend not only on
the IMF but also on stellar nucleosynthesis yields.

Combining (\ref{eq:dratio}) and (\ref{eq:zratio2}) we get a new connection
between deuterium mass fraction and metal mass fraction in the form of
\beq
\label{eq:dz2}
\frac{D(t)}{D_p} = 
  \frac{\alpha}{\alpha+R} \left[ 1 + 
  \frac{R}{\alpha} \left( 1-\frac{Z(t)}{Z_{\rm max}} \right)^{1+R/\alpha}\right] 
\eeq
We see that deuterium monotonically decreases as 
the metal fraction grows from zero.  As metal fraction
$\rightarrow Z_{\rm max}$, we see $D \rightarrow D_{\rm min}$, as expected 
(\ref{eq:dmin}).

Figure \ref{fig:DvsZ}
illustrates the metal mass fraction evolution for our model
(\ref{eq:dz2}) for a
variety of choices of $\alpha$ \footnote{The curves
also fix $R = 0.2$ (left panel) and $R = 0.3$ (right panel),
but (\ref{eq:dz2})
shows that the curves only depend on the ratio $\alpha/R$, which
sets the scaling for other choices of $R$.}.
Because our focus is on the present-day ISM, 
the metallicity is a fairly well-determined value,
with metal mass fraction being about $Z_{\rm ISM} \approx Z_\odot$.
With this fixed, one can view (\ref{eq:dz2})
as a relationship between deuterium fraction and
the stellar {\em metal yield} $Z_{\rm synt}$, a quantity much less
observationally accessible. Consequently,
we plot deuterium fractions calculated from the (\ref{eq:dz2}) as a function of 
stellar metal yield 
as solid (blue) curves for different choices of scaling factor $\alpha$ for
the infall rate. We see that, as expected, the higher the infall rate
of pristine gas is, the higher the allowed stellar yields have to be. 
The lower band corresponds to the ratio of deuterium mass fractions 
$D_{\rm ISM}/D_p$ that is not corrected for dust depletion effects,
and allows metal yields in the range of $1 \la Z_{\rm synt}/Z_{\rm ISM} \la 4$ for $R=0.2$ and 
$Z_{\rm synt}/Z_{\rm ISM} \la 4 $ for $R=0.3$, and infall rates that
can at best be as high as the star-formation rate $\alpha \approx 1$, but would also be, due to
the mentioned universality, indistinguishable. 
On the other hand, top band 
plots the high \cite{linsky} ISM deuterium, which
allows for even higher infall rates, and
an even wider range of metal yields $3 \la Z_{\rm synt}/Z_{\rm ISM} \la 10$. 
This means that yields of metals from high-mass stars have to be at least $Z_{\rm synt} \approx 3 Z_{\rm ISM}$
in order for the recent ISM deuterium abundance estimate \cite{linsky} to be
consistent with the present ISM metallicity. 

A valid question now is if such mean metal yields are reasonable;
this depends on both the IMF and stellar yields themselves. 
Taking now the supernova metal yield to be $Z_{\rm SN}=10 Z_{\odot}$ \cite{rauscher, ww}
we find the total mass averaged metal yield 
for $R=0.3$ to be $Z_{\rm synt}/Z_{\rm ISM}=Z_{\rm SN}X_{\rm SN}R_{\rm SN}/R \approx 4$.
Such yield is not
consistent with the upper limit of the D measurements without dust
corrections (Figure \ref{fig:DvsZ} bottom band), but is consistent with the top deuterium
band that corresponds to dust corrections. We also see that a standard simple estimate
does, within uncertainties, allow reasonable stellar metal yields as required by the observed range 
of deuterium abundances. 

Moreover, Figure \ref{fig:DvsZ}
gives reasonable metal yields for the required infall rates presented on
Figure \ref{fig:Dvsgasf}. Specifically, with a return fraction set to $R=0.2$,
a wide range of allowed infall rates $\alpha$ (Figure \ref{fig:Dvsgasf}) allows also
for a wide range of metal yields. However, even though this
return fraction together with the observed gas fraction would allow infall rate
as high as $\alpha \approx 1.0$ (as shown on Figure \ref{fig:Dvsgasf}), the required metal
yields would in that case have to be at least $Z_{\rm synt}/Z_{\rm ISM} \ga 5$,
which is difficult to unambiguously rule out but would 
strain current results for both nucleosynthesis and the IMF.
Thus, metal yields in this case provide an additional constraint and
tighten the range of allowed infall rates. 
On the other hand $R=0.3$ case and the 
required infall parameter $0.5 \la \alpha \la 1.0$. (Figure \ref{fig:Dvsgasf}) allows
a reasonable minimal metal yield $Z_{\rm synt}/Z_{\rm ISM} \ga 2$.

However, Figure \ref{fig:DvsZ} also demonstrates that metal yield on its own is
not constraining of the infall rate. 
Though curves for different values of $\alpha$ terminate at different
$(Z_{\rm max},D_{\rm min})$ values, this is a small effect
compared to the fact that they all converge to same values
with almost no difference in their shapes. 
Thus the only way that deuterium-metal connection
can constraint different infall rates is by setting the termination point,
i.e. by setting the highest infall rate allowed, while all curves
for different allowed infall rates would be essentially indistinguishable.
This universality
with respect to the infall proportionality rate $\alpha$ 
demonstrates that the observables $D$ and $Z$ are uniquely
correlated. However, other observables like gas fraction, as seen earlier, and G-dwarf distribution
\cite{wakker} are indeed sensitive to infall rates and are as such good
infall indicators, unlike deuterium abundance, which is an excellent
indicator of the nature of metal production, but, as we have shown,
is a poor infall indicator.

{}

\end{document}